\documentclass[twocolumn,english,prb,aps,superscriptaddress,amsmath,amssymb,floatfix,longbibliography]{revtex4-2}
\usepackage[T1]{fontenc}
\usepackage{verbatim}
\usepackage{amsmath}
\usepackage{amssymb}
\usepackage{graphicx}

\makeatletter
\usepackage{times}
\usepackage{textcomp}
\usepackage{epstopdf}
\usepackage{braket}
\usepackage{tikz}
\usepackage[colorlinks,linkcolor=blue,citecolor=blue]{hyperref}
\usepackage{amsfonts}
\usepackage{bbm}
\usepackage{bbold} 

\pdfpageheight\paperheight
\pdfpagewidth\paperwidth

\@ifundefined{textcolor}{}{%
 \definecolor{BLACK}{gray}{0}
 \definecolor{WHITE}{gray}{1}
 \definecolor{RED}{rgb}{1,0,0}
 \definecolor{GREEN}{rgb}{0,1,0}
 \definecolor{BLUE}{rgb}{0,0,1}
 \definecolor{CYAN}{cmyk}{1,0,0,0}
 \definecolor{MAGENTA}{cmyk}{0,1,0,0}
 \definecolor{YELLOW}{cmyk}{0,0,1,0}
}

\usepackage{xcolor}\usepackage{soul}
\newcommand{\appropto}{\mathrel{\vcenter{
  \offinterlineskip\halign{\hfil$##$\cr
    \propto\cr\noalign{\kern2pt}\sim\cr\noalign{\kern-2pt}}}}}

\def \Tr {\mathrm{Tr}}
\definecolor{blue}{rgb}{0,0,1}
\definecolor{red}{rgb}{1,0,0}
\definecolor{green}{rgb}{0,1,0}

\newcommand{\beq}{\begin{equation}}
\newcommand{\eeq}{\end{equation}}

\newcommand{\Corho}{C_{O}^{\rho}}
\newcommand{\Sorho}{S_{O}^{\rho}}
\newcommand{\SorhoF}[1]{S'^{\rho_\sigma(#1)}_O}
\newcommand{\Dos}{\mathrm{DoS}}

\definecolor{darkgreen}{rgb}{0.0, 0.2, 0.13}
\definecolor{deeplilac}{rgb}{0.6, 0.33, 0.73}

\newcommand{\identity}{1\kern-0.25em\text{l}}

\definecolor{lightblue}{rgb}{0.19,0.55,0.91}

\definecolor{darkorange}{rgb}{1.0, 0.55, 0.0}

\usepackage[normalem]{ulem}
\newcommand\redsout{\bgroup\markoverwith{\textcolor{red}{\rule[0.5ex]{2pt}{0.4pt}}}\ULon}

\usepackage{babel}
\begin{document}

\title{Probing Off-diagonal Eigenstate Thermalization with Tensor Networks}

\author{Maxine Luo}
\affiliation{Max-Planck-Institut f\"ur Quantenoptik, Hans-Kopfermann-Stra{\ss}e 1, D-85748 Garching, Germany}
\affiliation{Munich Center for Quantum Science and Technology, Schellingstra{\ss}e 4, 80799 M\"unchen, Germany}
\author{Rahul Trivedi}
\affiliation{Max-Planck-Institut f\"ur Quantenoptik, Hans-Kopfermann-Stra{\ss}e 1, D-85748 Garching, Germany}
\affiliation{Electrical and Computer Engineering, University of Washington, Seattle, Washington 98195, USA
}
\author{Mari Carmen Ba\~nuls}
\affiliation{Max-Planck-Institut f\"ur Quantenoptik, Hans-Kopfermann-Stra{\ss}e 1, D-85748 Garching, Germany}
\affiliation{Munich Center for Quantum Science and Technology, Schellingstra{\ss}e 4, 80799 M\"unchen, Germany}
\author{J. Ignacio Cirac}
\affiliation{Max-Planck-Institut f\"ur Quantenoptik, Hans-Kopfermann-Stra{\ss}e 1, D-85748 Garching, Germany}
\affiliation{Munich Center for Quantum Science and Technology, Schellingstra{\ss}e 4, 80799 M\"unchen, Germany}
\begin{abstract}

Energy filter methods in combination with quantum simulation can efficiently access the properties of quantum many-body systems at finite energy densities [Lu et al. PRX Quantum 2, 020321 (2021)]. 
Classically simulating this algorithm with tensor networks can be used to investigate the microcanonical properties of large spin chains, as recently shown in [Yang et al. Phys. Rev. B 106, 024307 (2022)].
Here we extend this strategy to explore the properties of off-diagonal matrix elements of observables in the energy eigenbasis, fundamentally connected to the thermalization behavior and the eigenstate thermalization hypothesis.
We test the method on integrable and non-integrable spin chains of up to 60 sites, much larger than accessible with exact diagonalization. 
Our results allow us to explore the scaling of the off-diagonal functions with the size and energy difference, and to establish quantitative differences between integrable and 
non-integrable cases. 

\end{abstract}

\maketitle

\section{introduction}

Since the early days of quantum mechanics, the emergence of thermalization behavior in isolated quantum systems has been a fundamental and intriguing question~
\cite{neumann1929beweis, goldstein2010long}.
But it has been only in recent years that, thanks to the high levels of control and isolation of ultracold atomic experiments, it has become possible to explore the quantum thermalization phenomenon experimentally~\cite{trotzky2012probing,kaufman2016quantum,clos2016time}, which has rekindled the attention to the topic.

A theoretical keystone to explain quantum thermalization is provided by the Eigenstate Thermalization Hypothesis (ETH) \cite{Deutsch91,MarkSrednicki_1999, Deutsch2018eth, ethreview,ther&pre}. Connecting quantum many-body systems with random matrix theory,
the ETH conjectures a generic form for the matrix elements of physical observables in the energy eigenbasis of the system. The ansatz is expected to apply for large generic (chaotic) systems \cite{ethreview, ther&pre, Reimann2008Foundation, Reimann_2015, nation2018off, Reimann2021Refining}, whereas it can be violated, for instance, in integrable models \cite{Rigol2007GGE, vidmar2016generalized, essler2013time, essler2016quench, essler2023statistics} and strongly disordered models \cite{mblreview, luitz2015many, luitz2017ergodic}. 
The validity of ETH has been numerically probed for a number of models \cite{rigol2008thermalization,Brenes2020ETHinLocallyPerturbed,Brenes2020low-frenquecyXXZ,autocorr,2dtfi,pushthelimit,testall,rigol2009breakdown,rigol2009quantum, LIOM, beugeling2015off-diagonal, LeBlond2019Entanglement, LeBlond2020breaksSymmetries,Zhang2022Statsitical,anomalousThermalization}. However, most numerical studies rely on exact diagonalization (ED), which becomes infeasible for large systems due to the exponential scaling of the Hilbert space dimension with the system size. Investigating the structure of matrix elements as a function of the latter, and finding their asymptotic behavior, remains numerically challenging.

Tensor network (TN) methods offer possibilities to numerically explore quantum systems of sizes much larger than the ones allowed by ED. The most successful TN methods target equilibrium states (ground or low energy eigenstates, or thermal equilibrium~\cite{Verstraete2008mpsreview,SCHOLLWOCK201196,PAECKEL2019167998}). But probing ETH requires investigating states at finite energy, which typically are not efficiently described by a TN ansatz~\cite{Bianchi2022volume}. Nevertheless, a new algorithm has been recently proposed that precisely allows studying an ensemble of eigenstates at finite energy density~\cite{Lu2021,Yang2022}. The method simulates the effect of a narrow energy filter operator through its expansion as a sum of evolution operators. By (quantum or classically) simulating each of these evolutions and post-processing the data, it is possible to approximate the properties of a microcanonical ensemble over an extensive region of the spectrum. Classically simulating the evolution with tensor networks imposes a limit on the width of the accessible filters, but, as demonstrated in~\cite{Yang2022} it suffices to efficiently access the microcanonical values for spin chains up to 80 sites, thus providing a way to probe the diagonal part of the ETH ansatz.

In this paper, we generalize the applications of the energy filter method to probe the more challenging off-diagonal part of ETH.
More concretely, our method computes a (broadened) filter spectral function for a given (local) operator. 
{This function can be interpreted as an average of matrix elements over eigenstate pairs selected by two spectral filters, one selecting the average energy and the other the energy difference.
We demonstrate how, by simulating finite time evolutions with standard tensor network routines, in the spirit of~\cite{Yang2022}, we can obtain the effect of two combined filters.}
Using our method we compute and compare the spectral functions for several Ising spin chains, including integrable and non-integrable clean systems, as well as a disordered one, for much larger system sizes than allowed by exact diagonalization.
For the region we can reliably probe, we obtain convergence of the spectral functions with system size, and are able to discriminate qualitatively distinct features, such as a different scaling with energy difference.

The rest of the paper is structured as follows. Section \ref{sec:basics} provides a brief review of ETH and the filter ensemble. In section \ref{sec:method} we present the method used to compute the spectral function numerically with TNS algorithms. Section \ref{sec:result} describes the three different spin models studied, and collects our numerical results. In particular our results capture the asymptotic behavior of off-diagonal matrix elements, and exhibit qualitative differences in the energy and energy difference dependence between the integrable and generic cases. We also study how in the latter case, the fluctuation dissipation relation is fulfilled by the filter ensemble for large enough systems.

\section{Background}
\label{sec:basics}

\subsection{Eigenstate Thermalization Hypothesis (ETH)}
Consider a many-body Hamiltonian with spectral decomposition $\hat{H}=\sum_\alpha E_\alpha \ket{\alpha}\bra{\alpha}$. Given a physical observable $\hat{O}$, ETH predicts that its matrix elements $O_{\alpha\beta} \equiv \bra{\alpha}\hat{O}\ket{\beta}$ in the energy eigenbasis obey the following form \cite{MarkSrednicki_1999,ethreview}
\begin{align}
O_{\alpha\beta}= O(\bar{E}) \delta_{\alpha\beta} +e^{-\frac{S(\bar{E})}{2}}f_O(\bar{E},\omega)R_{\alpha\beta} .
\label{equ:eth}
\end{align}
where we define the energy variables $\bar{E} \equiv (E_\alpha+E_\beta)/2$ and $\omega \equiv E_{\beta}-E_\alpha$. $R_{\alpha\beta}$ is a random variable with zero mean and unit variance. 
The thermodynamic entropy $S(E)$ can be defined as the logarithm of the number of available states in the microcanonical window, $\Dos(E) \Delta E$. In the literature it is nevertheless common to drop the dependence on the window width, which contributes only a small constant \cite{microentropy_1990}, and use as definition $S(E)=\ln \Dos(E)$.
$O(\bar{E})$ and $f_O(\bar{E},\omega)$ are smooth functions of their arguments. While Eq.~\ref{equ:eth} is probably the most common one for the ETH ansatz, other expressions exist that use a different entropy factor~\cite{ther&pre}, which results in a slightly different definition of $f_O(\bar{E},\omega)$. In our case, we write the off-diagonal term as $e^{-{[S(E_\alpha)+S(E_\beta)]}/4} f_O(\bar{E},\omega) R_{\alpha\beta}$.

ETH is a sufficient condition for quantum thermalization \cite{ethreview}: starting from a (sufficiently narrow in energy) out-of-equilibrium state, if~\eqref{equ:eth} is satisfied, the time-averaged expectation value of physical observables will relax to its {microcanonical average $O(E)$} in the limit of infinitely long time, with the fluctuations around the thermalization value controlled by the off-diagonal matrix elements.
Furthermore, it is believed that ETH holds for generic non-integrable systems and few-body operators, and multiple numerical studies have been conducted to verify its validity in such scenarios ~\cite{rigol2008thermalization, Brenes2020ETHinLocallyPerturbed, Brenes2020low-frenquecyXXZ, autocorr, 2dtfi, pushthelimit, testall}. Violations of ETH can be observed in integrable systems \cite{rigol2009breakdown, rigol2009quantum, LIOM, beugeling2015off-diagonal, LeBlond2019Entanglement, LeBlond2020breaksSymmetries, Zhang2022Statsitical}, as well as strongly disordered models \cite{mblreview, luitz2015many, anomalousThermalization,luitz2017ergodic}, due to their extensive number of (quasilocal) integrals of motions.

The off-diagonal structure function $|f(\bar{E},\omega)|^2$ appearing in the ETH ansatz is related to dynamic properties, and also determines the fluctuation-dissipation theorem (FDT) of nonequilibrium states. Recent works have focused on studying some of its properties in generic and non-generic systems.
The statistics of off-diagonal matrix elements in integrable spin chains were analyzed in~\cite{Brenes2020low-frenquecyXXZ, LeBlond2019Entanglement, LeBlond2020breaksSymmetries, Zhang2022Statsitical, essler2023statistics} and its dependence on energy difference $\omega$ and system size $N$ in~\cite{ethreview, 2dtfi, Brenes2020low-frenquecyXXZ, beugeling2015off-diagonal, Brenes2020ETHinLocallyPerturbed, LeBlond2019Entanglement, LeBlond2020breaksSymmetries, Zhang2022Statsitical, autocorr, anomalousThermalization}. 
On the other hand, the off-diagonal matrix elements in disordered models can display spectral properties deviating from the ETH prediction, as for instance shown in \cite{mblreview,nandkishore2014spectral,johri2015many}.
Finally, the standard ETH does not make explicit predictions for the correlation between matrix elements, which is nevertheless related to quantum chaos, and has been a focus of recent works ~\cite{eth&otoc,beyondstandard,deviationsfromrandom, otoc-fine,boundonchaos, Chan2019Butterfly}.


\subsection{The filter ensemble}

The most direct way to verify ETH numerically is to diagonalize the many-body Hamiltonian and analyze the exact energy eigenstates in a targeted energy window. This is, however, limited to small systems of the order of 20 spins, as the dimension of the Hilbert space increases exponentially with the size. 

A potential workaround is to study, instead of the properties of individual eigenstates, those of an ensemble that is narrow in energy. This strategy has been recently followed in~\cite{Yang2019,Lu2021,Yang2022} to investigate finite energy properties using the filter ensemble 
\begin{align}
    \hat{\rho}_\sigma(E)  := \frac{g_\sigma (E-\hat{H})}{\Tr \left[ g_\sigma(E-\hat{H}) \right]}, 
    \label{eq:rho_FE}
\end{align}
where $g_\sigma(x)$ is the Gaussian function,
\begin{align}
    g_\sigma(x)\equiv \frac{1}{\sqrt{2\pi}\sigma} \exp\left[-\frac{x^2}{2\sigma^2}\right].
    \label{eq:g_fun}
\end{align}

The filter ensemble $\hat{\rho}_\sigma(E)$ is diagonal in the energy eigenbasis, and it is centered at $E$, with $\sigma$ being the energy width. 
The expectation values for $\hat{\rho}_\sigma(E)$ will converge to the microcanonical ones in the limit of small $\sigma$,
\begin{align}
    \Tr[\hat{\rho}_\sigma(E) \hat{O}] \xrightarrow[\sigma \to 0]{}  O(E),
    \label{eq:microcanonicalvalue}
\end{align}
such that the filter ensemble can be used to probe the diagonal part of the ETH. In~\cite{Yang2022} it was argued that, in a system fulfilling ETH, a width $\sigma = o(\sqrt{N})$ should be sufficient for the l.h.s. of \eqref{eq:microcanonicalvalue} to converge to the microcanonical values for intensive quantities (see also \cite{dymarsky2017canonical}).

\section{The spectral function of filter ensembles}
\label{sec:method}

In this work, we are interested in the application of filter methods to probe the off-diagonal structure of the ETH. In particular, we aim at extracting information about the off-diagonal structure function $|f_O(E,\omega)|^2$.
The main quantity in our study will be the spectral function of the filter ensemble, defined as follows. The autocorrelation for some local observable $\hat{O}$ in a given state $\hat{\rho}$ can be computed as
\beq
\Corho(t)=\Tr \left[\hat{\rho} \hat{O}(t) \hat{O}^\dagger \right].
\label{eq:autocorr}
\eeq
Its Fourier transform yields the spectral function $\Sorho(\omega)=\frac{1}{2\pi}\int_{-\infty}^{\infty} dt e^{i\omega t} \Corho (t)$. 
If the state is diagonal in the energy basis, we denote its matrix elements $\bra{\alpha} \hat{\rho}  \ket{\alpha}$. This is the case for a single eigenstate or for the Gibbs ensemble, but also for the filter ensemble defined in Eq.~\eqref{eq:rho_FE}. In such case, the spectral function
can be written as
\beq
\Sorho(\omega) =
 \sum_{\alpha\beta} \bra{\alpha} \hat{\rho} \ket{\alpha}\left|O_{\alpha\beta}\right| ^2 \delta(\omega - E_\beta + E_\alpha),
    \label{eq:spectral_function_rho}
\eeq
i.e. $\Sorho(\omega) $ is an average over squared matrix elements {$O_{\alpha \beta}=\bra{\alpha}\hat{O}\ket{\beta}$} between energy eigenstates with fixed energy difference $\omega$, 
weighted by the probability of the eigenstates $\ket{\alpha}$ in the distribution defined by $\rho$.
In the following we focus on the filter ensemble $\hat{\rho}_{\sigma}(E)$, for which
the corresponding probabilities are $\bra{\alpha} \hat{\rho} \ket{\alpha} \propto g_{\sigma}(E-E_{\alpha})$, for the function $g_{\sigma}$ defined in Eq.~\eqref{eq:g_fun}, with the normalization factor specified in Eq.~\eqref{eq:rho_FE}.

\begin{figure}[t]
    \centering
    \includegraphics[width=0.8\linewidth]{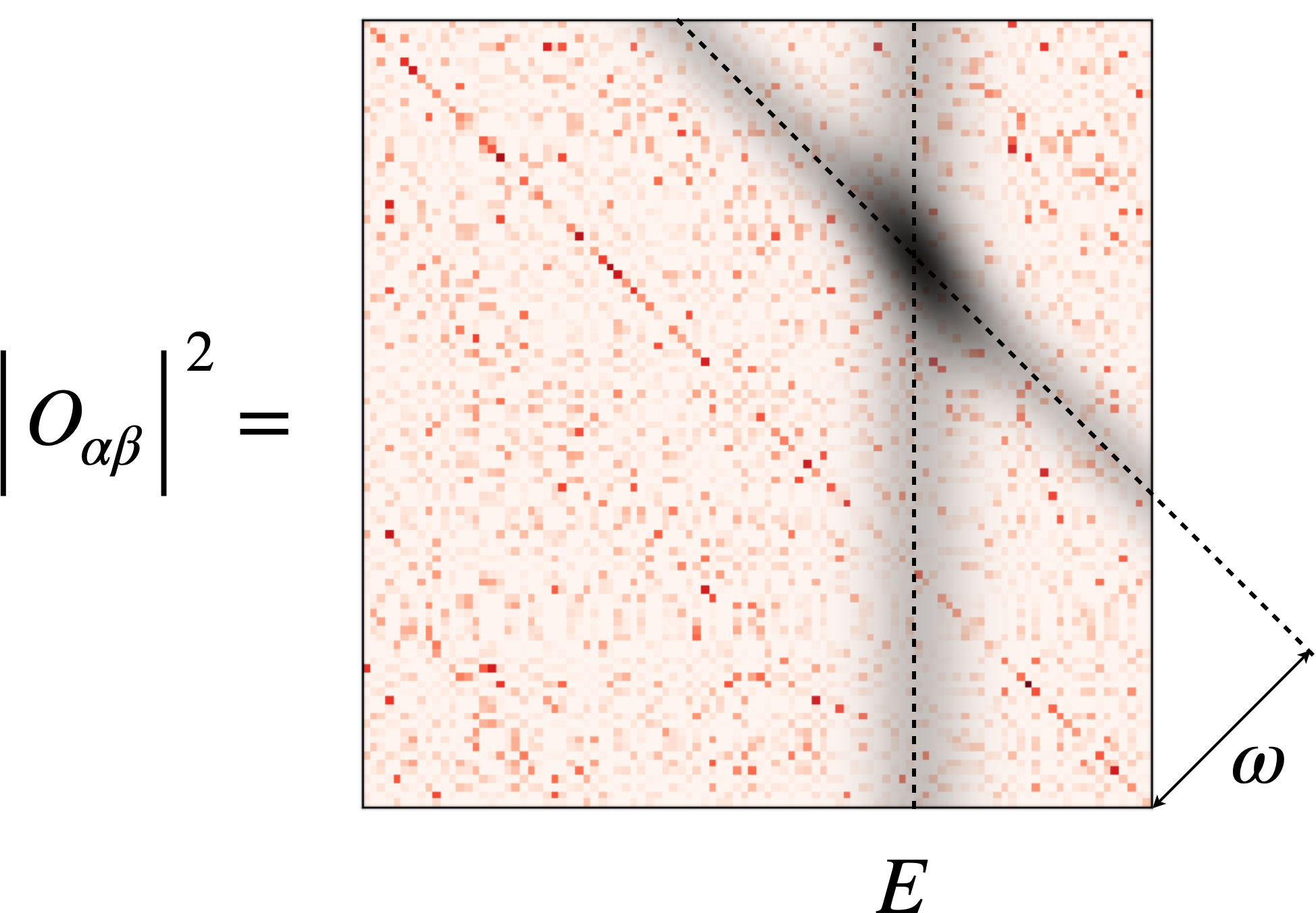}
    \caption{A graphical illustration of Eq. \ref{eq:s'_filter}. The heatmap shows the matrix elements $|O_{\alpha\beta}|^2$ in the eigenstate basis. The x-axis and y-axis are the eigenenergies $E_\alpha$ and $E_\beta$.
    The shadowed stripes indicate the filters on energy $E_\alpha$ and energy difference $E_\beta-E_\alpha$. The summation occurs over the matrix elements within the intersection of two shadowed stripes.
    }
    \label{fig:sketch}
\end{figure}

In order to compute this quantity, we introduce a generalized version of the spectral function, where we replace the 
$\delta$ function in Eq. \eqref{eq:spectral_function_rho} by a Gaussian of width $\sigma_{\omega}$, which we will approximate 
by a second filter acting on the energy difference $E_{\beta}-E_{\alpha}$.
This results in a broadened spectral function $ \SorhoF{E}(t)$ that 
performs an average of squared matrix elements $|O_{\alpha \beta}|^2$ for states $\alpha$ in the support of $\rho_{\sigma}(E)$ and states $\beta$ with energies around $E_{\alpha}+\omega$, 
as graphically illustrated in Fig.~\ref{fig:sketch}.
The function can be written as
\begin{align}
    \SorhoF{E}(\omega) 
        \equiv& \frac{\sum_{\alpha\beta}|O_{\alpha\beta}|^2 g_\sigma(E-E_\alpha) g_{\sigma_\omega}(\omega-E_\beta+E_\alpha)}{\sum_\eta g_\sigma(E-E_\eta)}.
        \label{eq:s'_filter}
\end{align}
For a local and bounded Hamiltonian, the density of states converges weakly to a Gaussian distribution in the thermodynamic limit, with a width of $\sqrt{N}\sigma_0$, where $\sigma_0$ is a constant independent of the system size \cite{hartmann2005spectral,keating2015spectra}. If the filters are narrow enough compared to $\sqrt{N}\sigma_0$, we can consider the density of states almost constant within the peak (while still much wider than the level spacing), and we can express $\SorhoF{E}$ in terms of an average of matrix elements as
\begin{align}
    S'^{\rho_\sigma\left(E\right)}_O(\omega) \approx e^{S(E+\omega)}
    \overline{|O_{E,E+\omega}|^2},
    \label{equ:averageofelements}
\end{align}
where the average is taken over pairs of eigenstates with energies around $E$ and $E+\omega$, respectively. 

\subsubsection{Relation with ETH}

Since the filter does not select precise energy differences, the sum will in general also pick up a contribution from the (much larger) diagonal matrix 
elements. 
In order to study the off-diagonal part of ETH, we thus choose observables for which the microcanonical value, and thus the diagonal contribution, vanishes, i.e. $O(E)=0$.
For these observables, the ETH ansatz predicts
\begin{equation}
\begin{aligned}
    \overline{|O_{E,E+\omega}|^2} 
    =& e^{-\frac{S(E+\omega)+S(E)}{2}} |f_O(E+\omega/2,\omega)|^2,
    \end{aligned}
\end{equation}
and thus
\beq
\SorhoF{E}(\omega) \approx e^{\frac{S(E+\omega)-S(E)}{2}} |f_O(E+\omega/2,\omega)|^2.
\label{equ:s_f}
\eeq
Therefore, $\SorhoF{E}(\omega)$ is directly related to the function $f_O$ and should also be a smooth function if ETH holds. 
The finite filter widths imply multiplicative corrections to Eq. \eqref{equ:s_f} of order $\mathcal{O}(\sigma^2/N^2+\sigma_\omega^2)$, as explicitly shown in appendix~\ref{app:sf}.

Since the filter strategy can be used to determine the density of states (see also~\cite{Yang2019,Papaefstathiou2021}), we can extract 
the value of $|f_O|$ from the computed spectral function.
But for Hermitian $\hat{O}$, a better strategy is 
making use of the fact that $f_O(E,\omega) = f_O(E,-\omega)$, which allows us to eliminate the exponential factor by combining the spectral functions at different arguments in 
a single function
\begin{equation}
    \begin{aligned}
    V_O(E,\omega) \equiv  \sqrt{\SorhoF{E-\omega/2}(\omega)  \SorhoF{E+\omega/2}(-\omega)} 
    {\approx} |f_O(E,\omega)|^2,
    \end{aligned}
    \label{eq:f_from_S}
\end{equation}
where the last part holds for ETH, as it follows from \eqref{equ:s_f}.
The function $V_O(E,\omega)$ can be understood as the variance of the matrix elements within the filter, and 
can obviously be computed for any system, fulfilling ETH or not, 
but in the latter case, it is not ensured to be a smooth function. 

Notice that our method is also applicable to operators with nonzero microcanonical values. In such case, the microcanonical value can be approximated using the method in \cite{Yang2019, Yang2022}, and subsequently subtracted from our calculations, to extract the off-diagonal component. 

Finally, it is worth noticing that the regularized correlators introduced in~\cite{pappalardi2024windows}, which can be related to filtered functions (see appendix~\ref{app:more_filters}), provide a similar strategy to extract the function $|f_O|$.

\subsubsection{The filter method}

Our numerical strategy is based on the TN simulation of the filter operators presented in~\cite{Yang2022}.
While the details are discussed in~\cite{banuls2019much, Faster_ground_state, Lu2021}, we sketch here the main steps for the sake of clarity. It is convenient to approximate the Gaussian by a cosine function as
\begin{align}
    \mathrm{e}^{-\xi^2/2 \sigma^2} \approx
     \cos^M\left(\xi/\alpha\right),
\end{align}
where $M=\lfloor(\alpha/\sigma)^2\rfloor_2$ ( $\lfloor\ldots \rfloor_2$ indicating the nearest even integer) and $\alpha$ is a rescaling factor introduced to ensure that the range of the argument $\xi/\alpha$ is smaller than the period of the cosine function $\pi$. 

Using the binomial expansion, the cosine power can be written as a sum of $M+1$ complex exponentials.
The number of terms in this sum can be reduced to $O(x \sqrt{M})$, by introducing a small error controlled by $x=O(1)$, yielding

\begin{align}
    g_\sigma(\xi) \approx \frac{1}{\alpha \pi c_0^{(M)}} \sum_{m=-x\sqrt{M}}^{x\sqrt{M}} c_m^ {(M)} e^{-i \xi t_m},
    \label{eq:Psigma}
\end{align}
where $t_m = 2m/\alpha$ and $c_m^{(M)} = \binom{M}{M/2 -m}/{2^M}$. 

Using the expansion~\eqref{eq:Psigma} for both Gaussian functions in Eq.~\eqref{eq:s'_filter} we obtain the following expression for the generalized spectral function $\SorhoF{E}(\omega)$ 
\begin{widetext}
\begin{equation}
\begin{aligned}
    \SorhoF{E}(\omega)
    = & \frac
{\sum_{m,n} c_m^{(M)} c_n^{(M_{\omega})} \sum_{\alpha\beta} \left| O_{\alpha\beta}\right|^2 e^{-i(E-E_\alpha)t_m+i(\omega-E_\beta + E_\alpha) t_n}   }
{ \alpha\pi c_0^{(M_{\omega})} \sum_{m} c_m^{(M)} \sum_{\eta}  e^{-i(E-E_\eta)t_m}}
    \\= &\frac
{\sum_{m,n} c_m^{(M)} c_n^{(M_{\omega})} e^{-iEt_m+i\omega t_n} \Tr [e^{iHt_m} \hat{O} (t_n)\hat{O}^\dagger]  }
{ \alpha\pi c_0^{(M_{\omega})} \sum_m c_m^{(M)} e^{-iEt_m} \Tr[e^{iHt_m}]},
\end{aligned}
\label{eq:doublesum}
\end{equation}
\end{widetext}
where in the second line we have simply used the trace and product of operators to rewrite the sums over the spectrum. Notice that both Gaussian filters have independent widths, corresponding to two different expansion parameters $M$ and $M_{\omega}$. Also the normalization factor $\alpha$ could in principle be different for each filter. Nevertheless, it is convenient for the numerics to use a common value of $\alpha$, so we choose the largest of both.

Notice that a different pair of filters could be used resulting in an average of matrix elements that probes the structure of the off-diagonal matrix elements using these filters. We briefly introduce other possibilities in App. \ref{app:num}.

\subsubsection{Cosine filter parameters}

The cosine filter for the ensemble is determined by the three parameters $(\sigma, \alpha, x)$. Similarly, the triplet  $(\sigma_\omega, \alpha_\omega, x_\omega)$ determines the 
cosine filter approximating the second Gaussian function in Eq.~\eqref{eq:s'_filter}.
The flexibility in parameter selection provides greater control over the performance of the approximation.

As mentioned above, for simplicity, we choose the same rescaling factor for both of the filters.  In practice we find $\alpha = \alpha_\omega> E_{max}-E_{min}$, with $E_{\max}$ ($E_{\min}$) being the highest (lowest) energy in the spectrum of $H$, to be enough to ensure the proper bound of both cosine filter arguments in the regime we study. In order to determine the suitable value, we estimate $E_{\max}$ ($E_{\min}$) using a variational MPS optimization, and choose $\alpha>1.1(E_{\max}-E_{\min})/\pi$. This fixes the time step in the filter expansion $\Delta t=2/\alpha$ to be the same for both filters, such that we can use common values for $t_m$ and $t_n$ in Eq.~\eqref{eq:doublesum}.

The longest time in the filter expansion scales as
\begin{align}
    t_{\max} = \frac{2x\sqrt{M}}{\alpha} = \frac{2x}{\sigma}.
    \label{eq:txsigma}
\end{align}
The smallest accessible widths are fixed by the longest times that we can reliably simulate using the TNS algorithms given the available computational resources and the finite precision of the numerical estimates.
In App. \ref{app:num} we discuss in detail how we choose $t_{\max}$ and $x$ for the specific models under study, to achieve the smallest filter widths while keeping the numerical error under control. Based on our error analysis results, we employ filter widths of $\sigma = \mathcal{O}(\sqrt{N})$ and $\sigma_\omega = \mathcal{O}(1)$. 

\subsubsection{Tensor network simulations}

Each of the terms in Eq.~\eqref{eq:doublesum} can be evaluated numerically with TNS techniques similar to the ones employed in~\cite{Yang2022}.
In particular, here we need to compute the trace expressions $\Tr[e^{iHt_m}]$ and $\Tr [e^{iHt_m}\hat{O}(t_n) \hat{O}^\dagger]$. To calculate the latter, we approximate
$e^{iH(t_m+t_n)/2}\hat{O} e^{-iHt_n/2}$ using matrix product operators (MPO) \cite{Verstraete2008mpsreview,SCHOLLWOCK201196,PAECKEL2019167998}
{at times $t_{\ell}=2\ell/\alpha$ ($\ell=m,\,n$), for $-x\sqrt{M}\leq m \leq x\sqrt{M}$ and $-x\sqrt{M_{\omega}}\leq n \leq x\sqrt{M_{\omega}}$}. In our simulation this is achieved by the time-evolving block decimation (TEBD) algorithm \cite{Verstraete2008mpsreview,SCHOLLWOCK201196,PAECKEL2019167998}. 
Evaluating the trace $\Tr [e^{iHt_m}\hat{O}(t_n) \hat{O}^\dagger]$ corresponds to computing the inner product of the two MPOs, which can be done efficiently. This strategy of evolving the MPO on both physical indices, splitting the evolution between both operators and evaluating an inner product has been shown to extend the time one can reach with a limited bond dimension~\cite{Karrasch2013finiteT,Barthel2013thermal}.

Similarly, in order to obtain $\Tr[e^{iHt_m}]$, we compute an MPO approximation to the evolution operator at times $t_m$. Evaluating the trace is then equivalent to a contraction with the vectorized identity operator.

We have studied several spin models for different system sizes, up to $N=60$. In all cases, we appoximated the evolution operators by a second order Trotter expansion with small Trotter step 
$0.01$. The results shown in the following were obtained with maximal bond dimension $D=600$, which
found to be enough to guarantee convergence for the studied timescales (see app.~\ref{app:num} for more detailed description of the numerical errors).

\section{Probing ETH with spectral functions: numerical results}
\label{sec:result}

\begin{figure*}[t]
    \centering
    \includegraphics[width=0.8\textwidth]{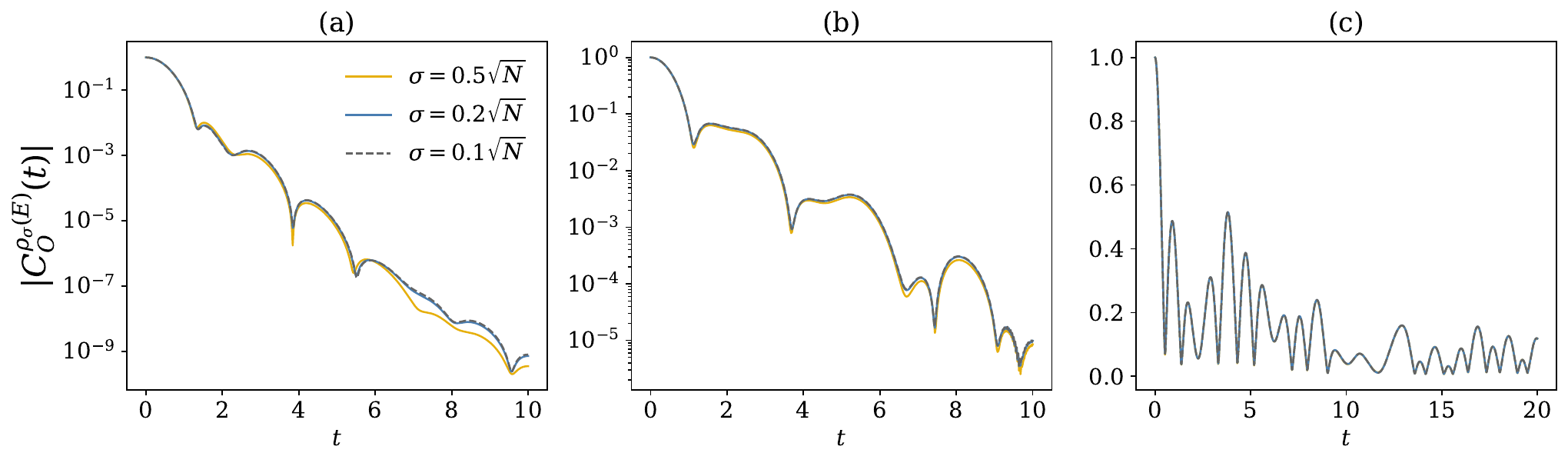}
    \caption{Absolute value of the autocorrelator $|C_O^{\rho_\sigma(E)}(t)|$ for $\hat{O}=\sigma_{N/2}^z$, as a function of time for different ensemble widths, 
    $\sigma=0.5\sqrt{N}$, $0.2\sqrt{N}$, $0.1\sqrt{N}$,  in a system of size $N=40$, at energy density $E/N=0.5$ for the integrable (a), non-integrable (b), and disordered system (c). }
    \label{fig:sigma}
\end{figure*}

\begin{figure*}[t]
    \centering
    \includegraphics[width=0.8\textwidth]{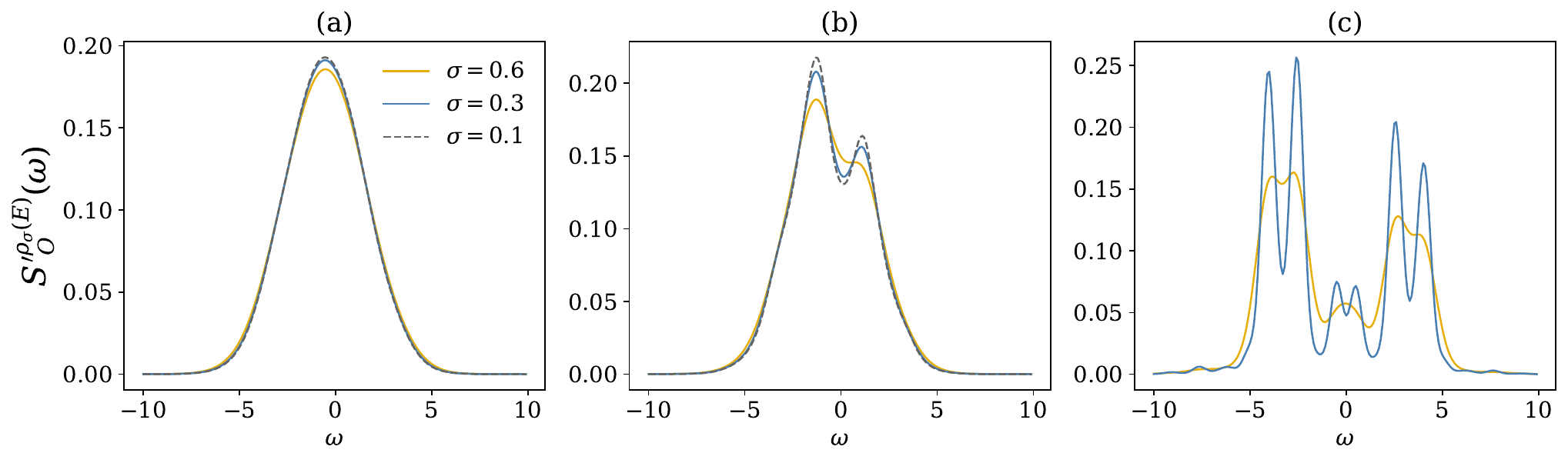}
    \caption{Generalized spectral functions $S'^{\rho_{\sigma}(E)}_O(\omega)$ for $\hat{O}=\sigma_{N/2}^z$, as a function of the energy difference $\omega$, 
    for a filter ensemble of width $\sigma=0.2\sqrt{N}$, at energy $E/N=0.5$, with $N=40$ and varying $\sigma_\omega=0.6$, $0.3$, $0.1$. The panels show the results for (a) integrable, (b) non-integrable, (c) disordered model. Notice that in the disordered system we impose $\sigma_\omega\ge 0.3$ to ensure the smallness of the numerical error (see App. \ref{app:num}). }
    \label{fig:sigma_omega}
\end{figure*}

\subsection{Setup} 
We benchmark the method on a quantum Ising chain with open boundary conditions, selectively including a disordered field and an integrability-breaking next-to-nearest-neighbor term,
\begin{align}
    \hat{H} = - J \sum_{i=1}^{N-1} \sigma_i^z \sigma_{i+1}^z - J_2 \sum_{i=1}^{N-2} \sigma_i^z \sigma_{i+2}^z -
    \sum_{i=1}^N (g+r_i) \sigma_i^x .
    \label{equ:ham}
\end{align}
$J$ sets the energy scale and in the following, we fix it to $J$=1. 
The model is integrable when $J_2=0$, {when it can be mapped to free fermions}. 
The transverse field includes a homogeneous component $h$, and potentially a disordered one $r_i$.
For simulations of the disordered model, the values of $r_i$ are sampled from the uniform distribution in the interval $[-r,r]$. 

We focus on three different sets of parameters. 
The first one, $(J_2,g,r)=(0.0,1.05,0)$ is the transverse field Ising model with uniform potential and thus integrable.
The second one is $(J_2,g,r)=(0.2,1.05,0)$, which corresponds to a non-integrable case, for which ETH is expected to hold. 
We finally consider also a disordered case, $(J_2,g,r)=(0.2,0,3.0)$, where the on-site field takes random values.

The observable we focus on is the longitudinal magnetization of the central site $\hat{O}=\sigma_{N/2}^z$.
Notice that $\hat{H}$ is invariant under the transformation $\mathcal{F}=\prod_{j=1}^{N}\sigma_j^x$, which flips all the spins in the chain. As $\sigma_{N/2}^z$ anti-commutes with $\mathcal{F}$, the expectation value of $\sigma_{N/2}^z$ in eigenstates of both $H$ and $\mathcal{F}$ is $0$, that is the diagonal elements are automatically zero and only off-diagonal elements contribute.
In the integrable system, where the eigenstates are characterized by free excitations, $\sigma_{N/2}^z$ is a many-particle operator in terms of these excitations, and thus the majority of off-diagonal elements are non-zero \cite{essler2023statistics}.

\begin{figure*}[t]
    \centering
    \includegraphics[width=0.8\textwidth]{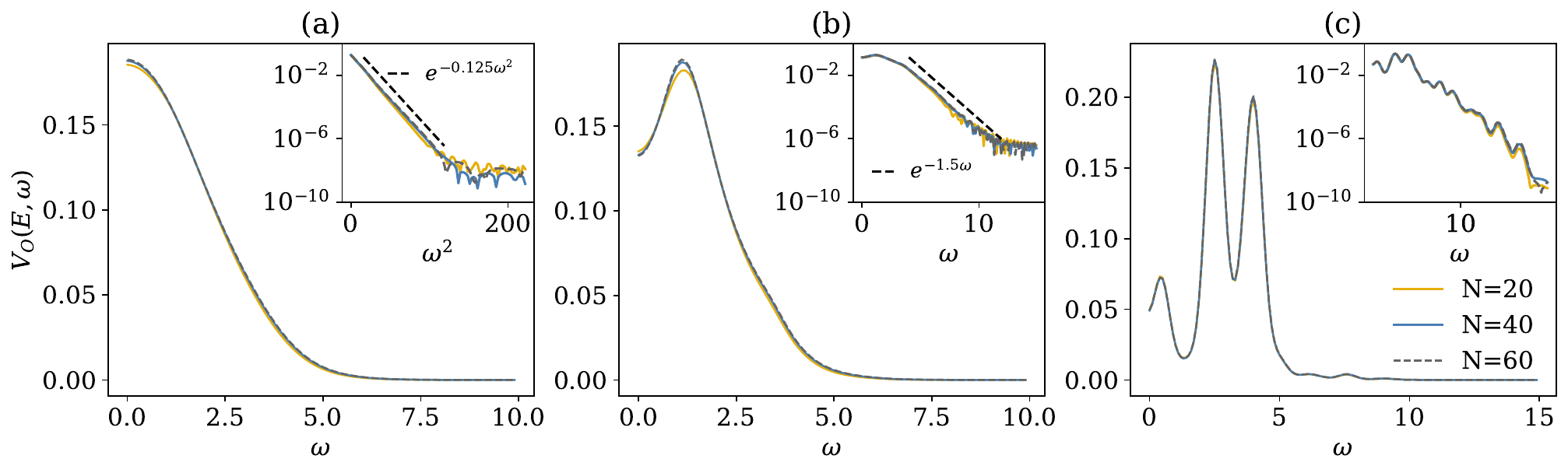}
    \caption{Size dependence of $V_O(E,\omega)$ for $\hat{O}=\sigma_{N/2}^z$, at fixed $E/N=0.5$, for $\sigma = 0.2 \sqrt{N}$, for system sizes $N=20$, 40, 60. The insets show $|V_O(E,\omega)|$ (in log-scale) as a function of $\omega$, with straight dashed lines {depicting fits to the functions $\propto e^{-k_1\omega^2}$ (a) and $\propto e^{-k_2\omega}$ (b).}  The different panels correspond to the various models: (a) integrable with $\sigma_\omega=0.1$, (b) non-integrable with $\sigma_\omega=0.1$, and (c) disordered system, for which $\sigma_\omega=0.3$.}
    \label{fig:size}
\end{figure*}

\subsection{Effect of the filter widths}

The spectral function obtained with our method depends on the filter parameters described above. Thus, first of all, we need to analyze the effect of the filter widths $\sigma$ and $\sigma_\omega$ on the results.
To isolate the influence of the filter ensemble width $\sigma$, we study the dependence of the autocorrelator $C_{O}^{\rho_{\sigma}}(t)$ (independent of $\sigma_{\omega}$) on this parameter. 
Figure~\ref{fig:sigma} shows, for each set of Hamiltonian parameters, the autocorrelator as a function of time for a chain of $N=40$ sites and an ensemble with mean energy density $E/N=0.5$, using filters of varying width $\sigma = q \sqrt{N}$, with proportionality factor $q=0.5$, $0.2$ and $0.1$.
We observe that the results are converged for $\sigma \leq 0.2 \sqrt{N}$. We observe a similar convergence for all the size and energy ranges studied in this work, thus 
 in the following we choose values of $\sigma$ within that range.

Additionally, from Fig.~\ref{fig:sigma} it becomes evident that the various studied models exhibit widely different time dependence of the autocorrelator.
In the clean systems (Figs. \ref{fig:sigma}(a) and \ref{fig:sigma}(b)), the autocorrelator decays exponentially with $t$ (notice the logarithmic scale of the vertical axis), whereas for the disordered system (Fig. \ref{fig:sigma}(c)) it remains significant at long times.

Next, to study the effect of the width of the energy-difference filter, 
we fix $\sigma = 0.2\sqrt{N}$ and vary $\sigma_\omega$.
Fig. \ref{fig:sigma_omega} shows, again for system size $N=40$, the generalized spectral function 
$\SorhoF{E}$ at $E/N=0.5$ for values
$\sigma_\omega = 0.6$, $0.3$, $0.1$.  
In the clean systems, the spectral function is a smooth function.
In contrast, for the disordered case, the function exhibits multiple peaks, at large values of the energy difference. This is a feature observed in localized systems~\cite{mblreview,nandkishore2014spectral,johri2015many}. 
The results shown in Fig.~\ref{fig:sigma_omega}(c) correspond to a particular realization of the disorder, but the figure is qualitatively similar for other realizations, with the positions of the peaks varying.

It is also worth noticing that in the disordered case the error originated by the sum truncation in \eqref{eq:Psigma} is more significant, so we need to choose a larger $x_{\omega}$ factor. We select 
$x_\omega\ge 3$ which, together with the upper bound on the simulated time $t_{\max}\leq 20$ imposed by the truncation error, means we can reach $\sigma_\omega\ge 0.3$ (see App. \ref{app:num}).

The previous arguments allow us to
fix the filter widths $\sigma$ and $\sigma_\omega$ to suitable values in the following studies. Specifically, we set the ensemble width to $\sigma=0.2\sqrt{N}$ for all systems. Thus, for the clean systems, we choose $\sigma_\omega=0.1$, and for the disordered model, $\sigma_\omega=0.3$.

\subsection{The off-diagonal matrix elements}

\begin{figure}[t]
    \centering
    \includegraphics[width=1.0\linewidth]{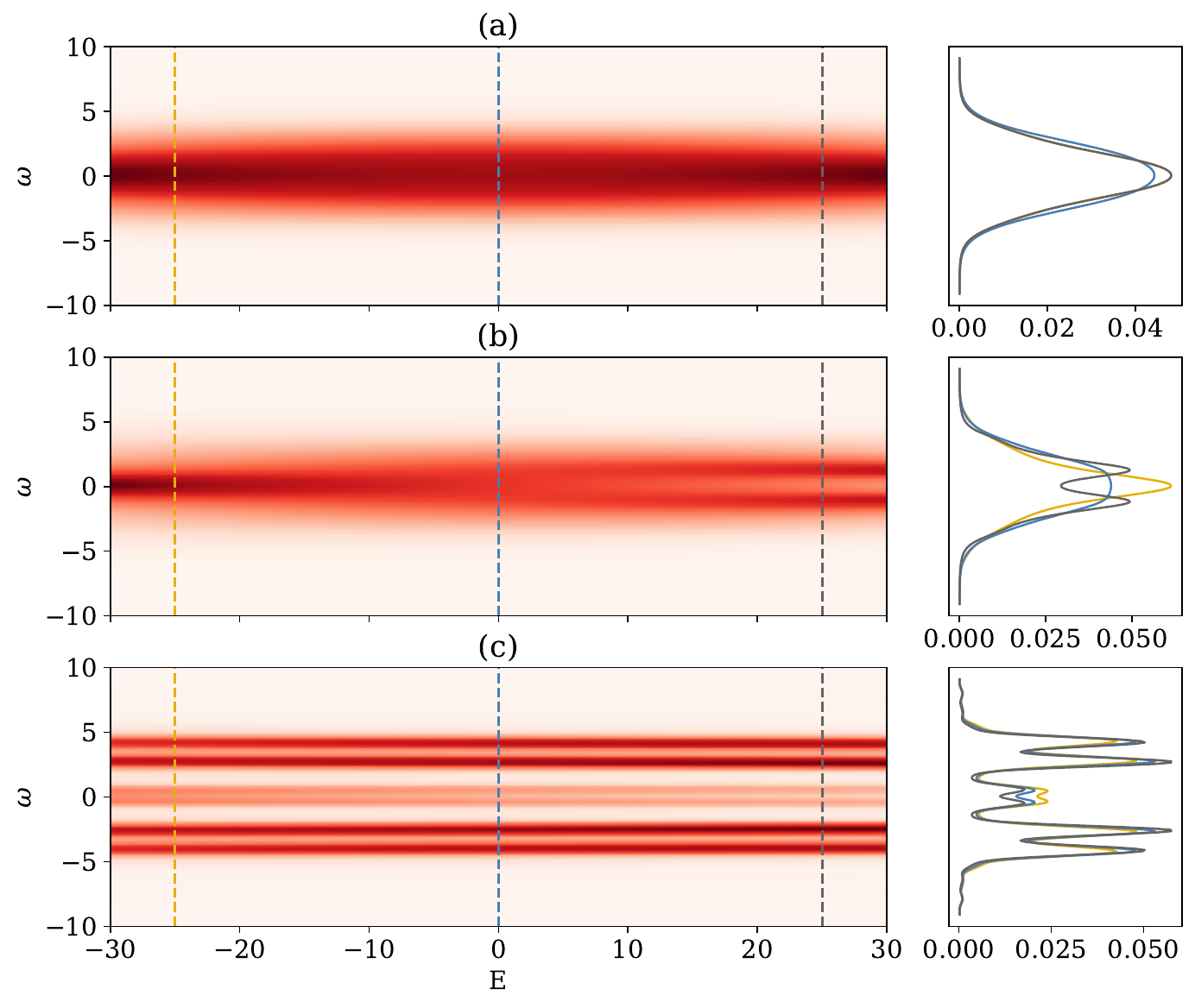}
    \caption{$V_O(E,\omega)$ for $\hat{O}=\sigma_{N/2}^z$, with $N=40$ and $\sigma = 0.2\sqrt{N}$. (a) Integrable system, $\sigma_\omega=0.1$. (b) Non-integrable system, $\sigma_\omega=0.1$. (c) Disordered system, $\sigma_\omega=0.3$. The plots on the right are profiles of the heatmaps at fixed mean energy density $E/N=0,\, \pm 5/8$.}
    \label{fig:f}
\end{figure}

To study the off-diagonal matrix elements, we analyze the function $V_O(E,\omega)$ in Eq. \eqref{eq:f_from_S}. {As discussed in section~\ref{sec:method},} $V_O(E,\omega)$ is the variance of the off-diagonal matrix elements within the filter, and it equals $|f_O(E,\omega)|^2$ if the system satisfies ETH. In Fig \ref{fig:f} we show the full $(E,\omega)$ dependence of $V_O(E,\omega)$ for the different studied Hamiltonian parameters, for system size $N=40$. 

Just as the spectral functions, $V_O(E,\omega)$ appears smooth in the {clean} systems [Fig. \ref{fig:f}(a,b)], whereas it exhibits multiple sharp peaks for the disordered system [Fig. \ref{fig:f}(c)].
This feature is also visible for the non-interacting ($J_2=0$) version of the same disordered chain and in the interacting case becomes more pronounced for stronger disorder, which seems to relate it to localization, even though the case shown in Fig. \ref{fig:f}(c) corresponds to moderate disorder strength.

Even though the function $V_O(E,\omega)$ is smooth for both clean cases,
significant differences are visible.
In the integrable system, $V_O(E,\omega)$ exhibits a symmetry $V_O(E,\omega) = V_O(-E,-\omega)$.
This property follows from the particle-hole symmetry in the integrable model, patent in its fermionic formulation, as explicitly shown in appendix~\ref{app:JW}.
In the non-integrable system there is no such symmetry, and $V_O(E,\omega)$ displays very different functional dependence on $\omega$ at high and low energies $E$, as seen in fig.~\ref{fig:f}(b).

A remarkable advantage of our method is that it can be applied to much larger systems than exact diagonalization, thus making it possible to address the system size scaling of these features. To study the system size dependence of $V_O(E,\omega)$, we plot the function in Fig. \ref{fig:size} at a fixed mean energy density $E/N=0.6$ for system sizes up to $N=60$, for the three models introduced above.
In all cases, we obtain convergence of the function with the system size, showing that the method allows us to observe the asymptotic behavior.
The figures show that for $\omega > 5$, $V_O(E,\omega)$ decreases very fast with $\omega$. However, the form of the decay is qualitatively different. As can be appreciated in the logarithmic plots shown in the insets, for the non-integrable case [Fig.~\ref{fig:size}(b)], the decay is exponential, as expected from ETH~\cite{ethreview}.
In contrast, for the integrable case, the decay of $V_O(E,\omega)$ is faster, compatible with $~\exp(-k_1\omega^2)$,
[see inset of Fig.~\ref{fig:size}(a)]. The $~\exp(-k_1\omega^2)$ decay of $V_O(E,\omega)$ at high frequency regime was also found in  \cite{LeBlond2019Entanglement,LeBlond2020breaksSymmetries,Zhang2022Statsitical} for other integrable models, such as XXZ chain and hard-core bosons.

\subsection{Numerical test of the Fluctuation-Dissipation Theorem}

The fluctuation-dissipation theorem is a general property of systems in thermodynamic equilibrium.
For quantum systems in the Gibbs state $\hat{\rho}_{\beta}= e^{-\beta \hat{H}}/\Tr[e^{-\beta \hat{H}}]$, the spectral function automatically satisfies the Kubo-Martin-Schwinger (KMS) condition~\cite{kubo1957statistical,martin1959theory}, which can be expressed
\begin{align}
    S^{\rho_\beta}_O(\omega) = e^{\beta \omega} S_O^{\rho_\beta} (-\omega),
\end{align}
This is a sufficient and necessary condition for the fluctuation-dissipation theorem (FDT) to hold. 

But the KMS condition can hold in more general situations. Beyond thermal equilibrium, it has been shown to hold in particular for some non-equilibrium initial states ~\cite{FDTafterQuench,et-fdr}.
More relevant for our case, it holds also for individual eigenstates~\cite{autocorr}, and for ensembles narrow in energy~\cite{ethreview, numFDT} in the thermodynamic limit when ETH is valid. We thus expect the filter ensemble to also fulfill FDT in the general case, at least in the limit of large systems. Using the filter strategy we can actually probe the validity of the relation in finite systems as a function of size. 

We can define the following indicator function that will test the KMS condition (hence the FDT) for a generic ensemble \cite{numFDT}, 
\begin{align}
    \beta_{\mathrm{FDT}}^{\rho}(\omega):= \frac{1}{\omega} \ln\left[\frac{S_O^{\rho}(\omega)}{S_O^{\rho}(-\omega)}\right].
    \label{equ:beta}
\end{align}
If the FDT holds we expect the function to be independent of $\omega$, and equal to the inverse microcanonical temperature corresponding to the mean energy of the ensemble $\rho$. 

For a generic system (fulfilling ETH), we can compute the value of \eqref{equ:beta} in the filter ensemble by expanding the spectral function around $E$ as
\begin{equation}
\begin{aligned}
    & S_O^{\rho_\sigma(E)}(\pm \omega) 
    \\ = & e^{\pm \frac{\beta \omega}{2} +\frac{3\omega^2}{8}\frac{\partial \beta}{\partial E}} \left(|f_O(E,\omega)|^2 \pm \frac{\omega}{2} \partial_E |f_O(E,\omega)|^2\right),
\end{aligned}
\end{equation}
where $\beta\equiv \partial_E S(E)$ is the inverse temperature at energy $E$ in the microcanonical ensemble. We obtain for the indicator function
\begin{equation}
\beta_{\mathrm{FDT}}^{\rho_\sigma(E)}(\omega)=\beta +\partial_E \ln \left| f_O(E,\omega)\right|^2.
\label{eq:beta_filter}
\end{equation}
We thus expect the function to be approximately equal to the inverse temperature, as KMS requires, with some correction. 
The major correction term $\partial_E \ln \left| f_O(E,\omega)\right|^2$ scales as $\mathcal{O}(1/N)$ (see App. \ref{app:sf} for more details). Thus we expect that, in the generic case, FDT indeed holds for the filter ensemble in the thermodynamic limit.  

\begin{figure}[t]
    \centering
    \includegraphics[width=1.0\linewidth]{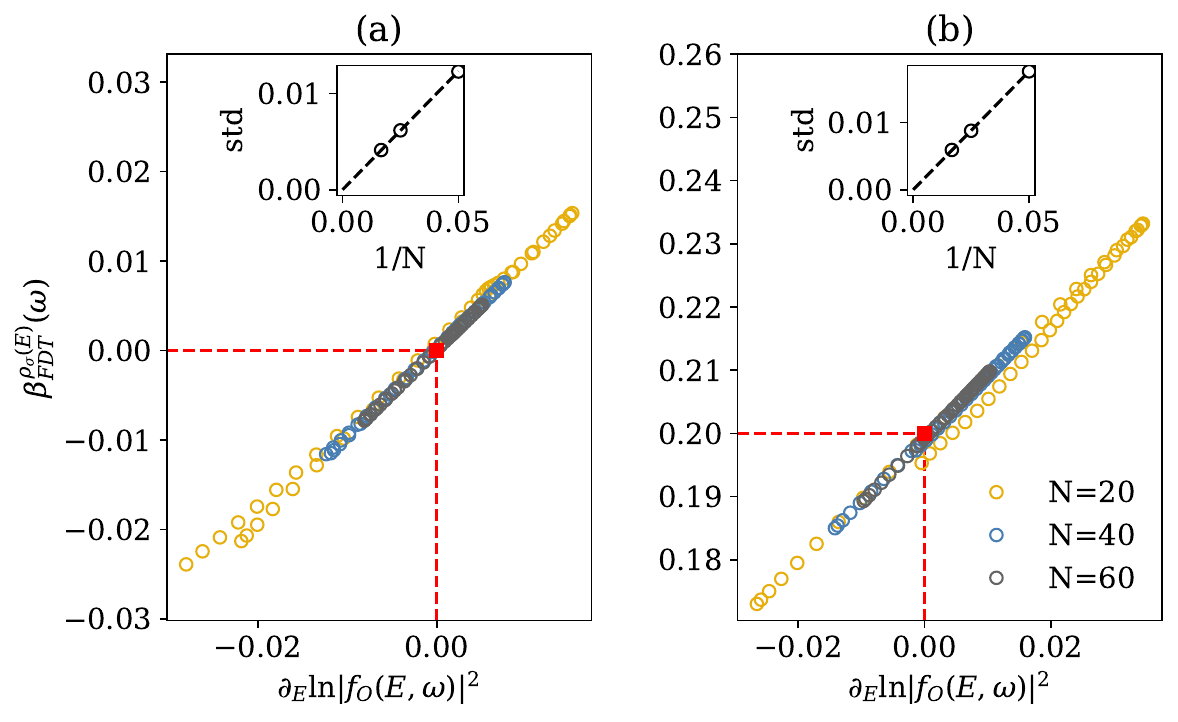}
    \caption{The FDT indicator functions $\beta_{\mathrm{FDT}}^{\rho_\sigma(E_0)}(\omega)$ v.s. $\partial_E \ln \left|f_O(E_0,\omega)\right|^2$ in the non-integrable system at sizes $N=20,40,60$, $\omega$ ranging from $[-5,5]$. $E_0$ corresponds to the reference inverse temperatures $\beta_0 = 0.0$ (a) and $0.2$ (b), marked in red. The observable is $\hat{O}=\sigma_{N/2}^z$. {Insets show the standard variance of $\beta_{\mathrm{FDT}}^{\rho_\sigma(E)}(\omega)$ over the interval $\omega\in [-5,5]$, scaling as $1/N$. }}
    \label{fig:FDT}
\end{figure}

We have computed this indicator function using our generalized spectral function for the 
non-integrable Ising chain, up to $N=60$ sites. 
To be able to compare different system sizes, we fix a value of the (reference) inverse temperature $\beta_0$. 
This corresponds to a value of the mean energy $E_0$ fulfilling $\beta_0 = \partial_E S(E_0)$, which we can determine from the results of the last section. To be specific, we obtain $\Dos(E)$ with filters and then extract $\partial_E S(E) = \partial_E \ln\Dos(E)$ using finite derivatives. 
{Then, we compute the indicator function Eq. \eqref{equ:beta} from the ratio between the values of the spectral function at $\omega$ and $-\omega$ at the corresponding energy $E_0$.}
To check the validity of our approach, we plot in Fig. \ref{fig:FDT} the value obtained for this indicator function as a function of $ \partial_E \ln \left|f_D(E_0,\omega)\right|^2$, which we obtained in an analogous way to $\partial_E S(E_0)$.
The figure shows this dependence for two values of the reference inverse temperature, $\beta_0=0$ and $0.2$, in the range $\omega\in [-5,5]$. 
The plots show a near-perfect linear dependence, {passing near the point $(0,\beta_0)$}, as predicted by Eq. \eqref{eq:beta_filter}. 

Our findings are consistent with those of the ED studies in \cite{numFDT, autocorr}, but extend the results to significantly larger system sizes.

\section{conclusions}

Using energy filter operators offers an alternative strategy to study the properties of the off-diagonal matrix elements of observables in the energy eigenbasis, and thus probe the ETH, fundamental ingredient in the theoretical understanding of quantum thermalization. 
In this work we have explored this possibility, with focus on the spectral function of the filter ensemble
$S_{O}^{\rho_\sigma(E)}(\omega)$.
We have shown that this quantity corresponds to an average of such matrix elements, and gives access to the off-diagonal function in the ETH ansatz, with corrections that depend on the filter width and the system size.

We have shown that this quantity can be simulated classically using TNS techniques, in a generalization of methods presented in~\cite{Yang2022} for the microcanonical averages.
In particular, this strategy allows addressing much larger systems than exact diagonalization, and allows us to observe convergence in the system size and thus to identify the asymptotic features of the off-diagonal matrix elements. 
It thus provides a powerful tool to explore the ETH ansatz.

In order to test the strategy, we have applied it to several operators in Ising chains including different terms, such that they span from integrable, non-integrable generic (ETH) to ergodicity breaking behaviors. In particular, we have shown that the spectral functions $S_{O}^{\rho_\sigma(E)}(\omega)$ for a non-integrable, generic instance and for a disordered chain exhibit clear qualitative and quantitative differences.

We have also shown that the filter spectral function provides a way to probe the validity of the FDT for the filter ensemble. In the limit of vanishing filter width this would converge to a probe of the relation for energy eigenstates, so far realized with ED for small systems~\cite{numFDT}.
Our numerical results show good agreement with FDT in the generic non-integrable chain up to 60 sites.

\acknowledgments
We are thankful to F. Essler for insightful discussions and suggesting the ANNI model example. We thank Yilun Yang for inspiring discussions and help in setting up simulations. This work was partially supported by the Deutsche Forschungsgemeinschaft (DFG, German Research Foundation) under Germany's Excellence Strategy -- EXC-2111 -- 390814868; SFB-TRR360
and by the EU-QUANTERA project TNiSQ (BA 6059/1-1).

\addcontentsline{toc}{section}{\refname}
\bibliography{method,eth,exp,analytical}

\appendix
\onecolumngrid

\section{Details in the filter method}


\label{app:num}
This appendix explains the filter method in a more compact operator language than in the previous works ~\cite{Lu2021, Yang2022}. We could define the following filter operators
\begin{equation}
    \begin{aligned}
        &\hat{P}_\sigma (E) \equiv g_\sigma\left[E-\hat{H}\right],
        \\&\hat{P}^c_\sigma (\omega) \equiv g_\sigma\left[\omega-\hat H\otimes \identity + \identity \otimes \hat{H}\right],
        \\& \hat{P}^a_\sigma (E) \equiv g_\sigma\left[E- \frac{\hat H\otimes \identity + \identity \otimes \hat{H}}{2}\right].
    \end{aligned}
\end{equation}
These are operators filtering, respectively, the energy value, the difference between two energy eigenvalues and the average energy of the pair. Notice that, while the first filter is an operator acting on the Hilbert space of the system, the last two are superoperators acting on operators. To describe them in a unified manner, we have used the operator-vector correspondence, in which each operator $O$ is mapped to a vector $|O)$ by mapping the basis elements $\ket{i}\bra{j}\to \ket{i}\otimes\ket{j}$, with the Hilbert-Schmidt inner product $(O_1|O_2) = \Tr[\hat{O}_1^\dagger\hat{O}_2]$. 
In this language,  $\hat{X}\otimes\hat{Y}|O) = |XOY^T )$.

Using this operator notation, the generalized spectral function defined in \eqref{eq:s'_filter} can be expressed as
\begin{align}
    S'^{\rho_\sigma(E)}_O(\omega)  =& \frac{( O | (\hat{P}_\sigma(E) \otimes \identity ) \hat{P}^c _{\sigma_\omega} (\omega) | O )}{ \Tr[\hat{P}_\sigma(E)]} \equiv \frac{A(E,\omega)}{B(E)},
    \label{eq:s'inoperaotr}
\end{align}
where in the last equality we have defined  
$A(E,\omega)$ and $B(E)$ as the numerator and the denominator of $S'^{\rho_\sigma(E)}_O(\omega)$.

As in Eq. \eqref{eq:Psigma}, each of these Gaussian filters can be approximated by a cosine filter, and truncated to a sum of exponentials of the argument, which correspond to time evolution operators. 
In the case of $\hat{P}_\sigma(E)$, these are regular evolution operators, generated by the Hamiltonian of the system, whereas for $\hat{P}^c _{\sigma_\omega}(\omega)$, they are superoperators generated by the commutator $H\otimes \identity - \identity \otimes \hat{H}$. 
Writing the sums explicitly, we obtain 
the following expressions for $A(E,\omega)$ and $B(E)$,
\begin{align}
     &A(E,\omega) = \frac{1}{\alpha^2\pi^2 c_0^{(M)} c_0^{(M_\omega)}}
{\sum_{m,n} c_m^{(M)} c_n^{(M_{\omega})} e^{-iEt_m+i\omega t_n} \Tr [e^{i\hat{H}(t_m+t_n)} \hat{O} e^{-i\hat{H}t_n}\hat{O}^\dagger]  } \nonumber \\
&\phantom{A(E,\omega) }
=\frac{1}{\alpha^2\pi^2 c_0^{(M)} c_0^{(M_\omega)}}
{\sum_{m,n} c_m^{(M)} c_n^{(M_{\omega})} e^{-iEt_m+i\omega t_n} \Tr [e^{i\hat{H}t_m}\hat{O}(t_n)\hat{O}^\dagger]  }
,
\\ &B(E) = \frac{1}{ \alpha\pi c_0^{(M)} }{\sum_m c_m^{(M)} e^{-iEt_m} \Tr[e^{iHt_m}]}.
\end{align}
$A(E,\omega)/B(E)$ then results in the expression Eq. \eqref{eq:doublesum}, which can be computed in practice.

The cosine filter approximation of $\hat{P}_\sigma (E) $ is determined by the three parameters $(\sigma, \alpha, x)$, while $(\sigma_\omega, \alpha_\omega, x_\omega)$ determines $\hat{P}^c _{\sigma_\omega} (\omega)$. 
The next paragrahs detail how we choose the values of the parameters to keep the errors under control.

\subsection{The rescaling factor $\alpha$}

The cosine filter has a period $\alpha\pi$. For the validity of the cosine filter scheme, its period should be larger than the regime we study. For the filter operator on energy $\hat{P}_\sigma(E)$, $\alpha \pi > E_{max}-E_{min}$ is required. While for the filter on energy difference $\hat{P}^c _{\sigma_\omega} (\omega)$, from Fig. \ref{fig:size} we find that at large $\omega$, $|V_O(E,\omega)|$ is smaller than the numerical errors. $\alpha_\omega \pi  > 2 \times 20$ is enough to cover the small $\omega$ regime that we are interested in. As mentioned in the main text, we choose the same rescaling factor for both of the filters for simplicity. $\alpha = \alpha_\omega> E_{max}-E_{min}$ would suffices in our studies.

\subsection{The factors $\sigma,x$ and numerical errors}
To analyze the error, it is convenient to look at $A(E,\omega)$ and $B(E)$ separately. When we replace the Gaussian filter operators with truncated sums, the dropped terms will introduce to the denominator $B(E)$ an error of
\begin{align}
    & \epsilon_{B} \leq  \frac{2}{\alpha\pi c_0^{(M)}} \sum_{m=x\sqrt{M}}^{M} c_m^{(M)} \left|\Tr[e^{iHt_m}]\right| \lesssim \frac{ \max(t_m)}{\sqrt{2\pi}x} \sum_{m=x\sqrt{M}}^{M} c_m^{(M)} \left|\Tr[e^{iHt_m}]\right|,
    \label{eq:epsilonB}
\end{align}
Where we have used $\frac{1}{\alpha \pi c_0^{M}} \approx \frac{1}{\sqrt{2\pi} \sigma} = \frac{\max(t_m)}{2\sqrt{2\pi} x }$. Similarly, the error in the numerator is
\begin{equation}
\begin{aligned}
    \epsilon_{A} \leq & \frac{2}{\alpha^2\pi^2 c_0^{(M)}c_0^{(M_\omega)}} 
    \left[ \sum_{m=x\sqrt{M}}^{M} \sum_{n=-M_\omega}^{M_\omega} + 
  \sum_{m=-M}^M\sum_{n=x_\omega\sqrt{M_\omega}}^{M_\omega} \right] 
    c_m^{(M)} c_n^{(M_\omega)} \left| \Tr [e^{iHt_m} \hat{O} (t_n)\hat{O}^\dagger]\right|
\end{aligned}
\end{equation}
To simplify this expression, we estimate the scale of $\left|\Tr [e^{iHt_m} \hat{O} (t_n)\hat{O}^\dagger] \right|$. Notice that it can be written as
\begin{align}
    \left|\Tr [e^{iHt_m} \hat{O} (t_n)\hat{O}^\dagger] \right| = \left| \sum_{\alpha \beta} e^{iE_\alpha t_m + i (E_\alpha-E_\beta) t_n} |\bra{\alpha}O\ket{\beta}|^2 \right|.
\end{align}
When $|t_n|$ or $|t_m|$ increases, the phases of the terms become less aligned, which in general results in a reduction of the summation. Therefore we expect
\begin{align}
    & \left|\Tr [e^{iHt_m} \hat{O} (t_n)\hat{O}^\dagger] \right| \lesssim \left|\Tr [ \hat{O} (t_n)\hat{O}^\dagger] \right|, \label{eq:a8}
    \\ & \left|\Tr [e^{iHt_m} \hat{O} (t_n)\hat{O}^\dagger] \right|  \lesssim \left|\Tr [e^{iHt_m} \hat{O}\hat{O}^\dagger] \right|, \label{eq:a9}
\end{align}
which we find always hold in our simulation. Using Eq. \eqref{eq:a8} and \eqref{eq:a9}, we have
\begin{equation}
\begin{aligned}
    & \epsilon_{A} \lesssim &  \frac{ \max(t_m) \max(t_n)}{ 4\pi x x_\omega}  \left[\sum_{m=x\sqrt{M}}^{M} c_m^{(M)} \left|\Tr [e^{iHt_m}\hat{O}\hat{O}^\dagger] \right| 
    +  \sum_{n=x_\omega\sqrt{M_\omega}}^{M_\omega} c_n^{(M_{\omega})} \left|\Tr [\hat{O} (t_n)\hat{O}^\dagger]\right| \right]. 
    \label{eq:epsilonA}
\end{aligned}    
\end{equation}
where $\sum_{m=-M}^{M} c_m^{(M)}=1$ is used. Eq. \eqref{eq:epsilonB} and \eqref{eq:epsilonA} show that the truncation error is negatively related to $x$. In the following, we choose $x$ as a $O(1)$ parameter in system size and adjust it for the models considered in the main text.

\subsubsection{$x$ and $\sigma$}
\paragraph{The error $\epsilon_B$ and the factor $x$}
To analyze the error $\epsilon_B$, we plot $|\Tr [e^{iHt_m }]|$ for the models and the operator in Section \ref{sec:result}. It shows $|\Tr [e^{iHt_m }]|$ decays exponentially fast with $t_m$, so that, after a finite time, the values become too small to distinguish them from zero, for any fixed numerical precision. We thus fix the largest simulated $t_m$ as the value when $\Tr [e^{iHt_m }]/\Tr [\identity]$ falls below a fixed threshold $10^{-5}$.

\begin{figure}[htbp]
    \centering
    \includegraphics[width=0.9\textwidth]{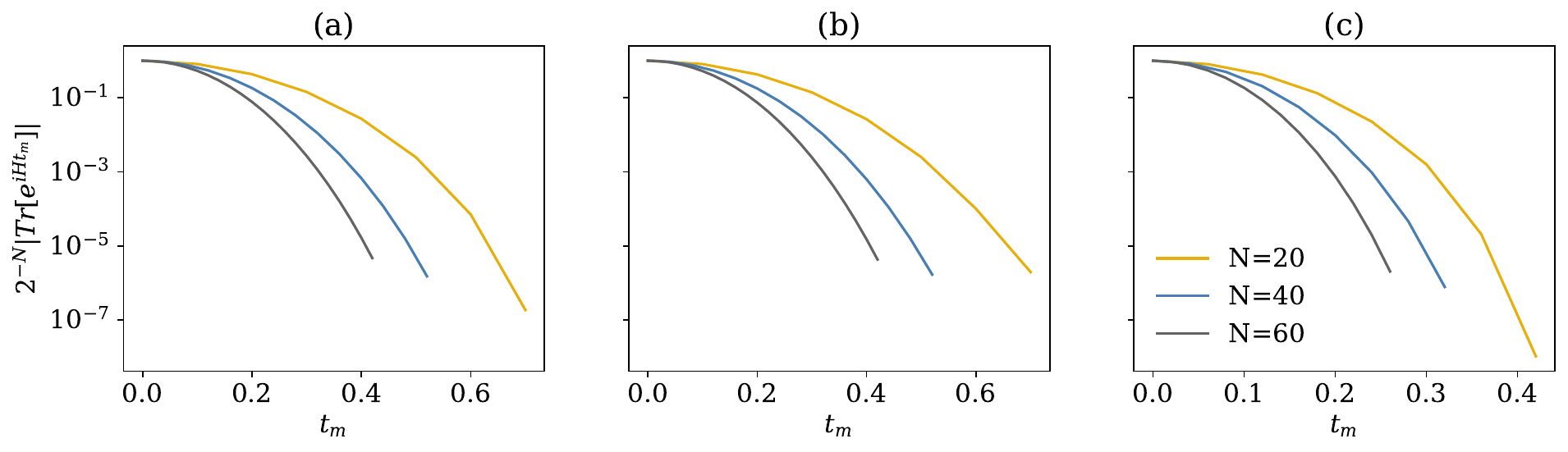}
    \caption{$2^{-N}|\Tr[e^{iHt_m}]|$ as a function of $t_m$. The simulation is terminated when $2^{-N}\Tr[e^{iHt_m}]$ falls below $10^{-5}$. (a) The intergrable system. (b) The Non-integrable system. (c) The disordered system.}
    \label{fig:cutoff1}
\end{figure}

Until this short time simulation, the MPO truncation error is small, 
and thus we can assume that the error comes from the truncation of the sum in $m$.
From Fig. \ref{fig:cutoff1} we assume that the truncated terms of $\Tr [e^{iHt_m }]$ are smaller than $10^{-5}\times 2^N$. According to Eq. \eqref{eq:epsilonB}, the error in the denominator $B(E)$ can be upper bounded as
\begin{align}
    \epsilon_B  \leq \frac{ 1} {\sqrt{2\pi} x }e^{-x^2/2} \times 10^{-5}\times 2^N,
\end{align}
where we used $\sum_{m=x\sqrt{M}}^{M}c_m^{(M)}\leq e^{-x^2/2}$.
Compared to $B(E)\approx \Dos(E)$ itself, we find that the relative error is small for any fixed $x\sim 1$. 

\paragraph{The filter width $\sigma$}
For fixed $x$, the filter width $\sigma$ inversely depends on the $\max(t_m)$.
As observed from Figure \ref{fig:cutoff1}, the cutoff time $\max(t_m)$ becomes shorter for larger systems. Here we provide a theoretical explanation of the dependence of $\max(t_m)$ on system sizes. For a traceless, local and bounded Hamiltonian, the density of states $\Dos(E)$ converges weakly to Gaussian distribution \cite{hartmann2005spectral,keating2015spectra} in the thermodynamic limit with width proportional to $\sqrt{N}$:
\begin{align}
    \int_{-\infty}^{E_0} \Dos(E) dE \xrightarrow{N\to \infty} \int _{-\infty} ^{E_0} \frac{d^Ne^{-E^2/2N\sigma_0^2}}{\sqrt{2\pi N}\sigma_0} dE
    \label{equ:dos}
\end{align}
where $d$ is the local Hilbert space and $\sigma_0$ is a constant independent of the system size. $\Tr[e^{iHt}]=\int d E e^{iEt} \Dos(E)$ is the Fourier transform of $\Dos(E)$ into the time domain. Therefore $\Tr[e^{iHt}]$ is also Gaussian, with width proportional to $1/\sqrt{N}$. As a result, the time for $\Tr[e^{iHt}]$ to fall below $10^{-5}$ of the initial value also scales as $\mathcal{O}(1/\sqrt{N})$.
According to Eq.~\eqref{eq:txsigma}, this corresponds to a filter width $\sigma=\mathcal{O}(\sqrt{N})$.

\subsubsection{$x_\omega$ and $\sigma_\omega$}
To analyze the error $\epsilon_A$, we need to study the time dependence of $|\Tr [e^{iHt_m}\hat{O}\hat{O}^\dagger] |$ and $|\Tr [\hat{O}(t_n) \hat{O}^\dagger]|$. For the observable $\hat{O}=\sigma_{N/2}^z$ we studied in the main text, $|\Tr [e^{iHt_m}\hat{O}\hat{O}^\dagger]|$ is simply $\left|\Tr [e^{iHt_m}] \right|$, and therefore the argument above also applies here. As for 
the time dependence of $|\Tr [\hat{O}(t_n) \hat{O}^\dagger]|$, we find it varies for different models. We thus simulate values of $t_n$ as large as possible, until the quantity becomes too small, or the truncation error in our MPO approximation becomes significant.

\begin{figure}[h]
    \centering
    \includegraphics[width=0.9\textwidth]{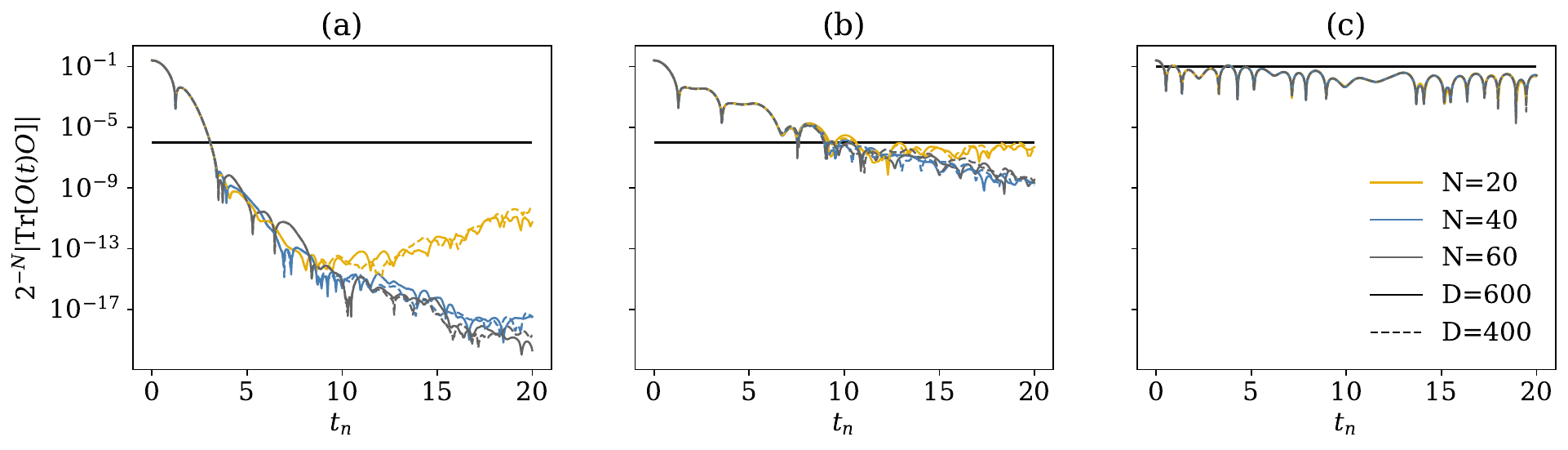}
    \caption{MPO simulation results of $|\Tr[O(t_n)O]|$ as a function of $t_n$, for both systems, system sizes $N=20,40,60$ and bond dimension $D=400,600$. The black lines indicate $10^{-6}$ or $10^{-1}$. (a) The intergrable system. (b) The Non-integrable system. (c) The disordered system.}
    \label{fig:error}
\end{figure}

In Fig. \ref{fig:error}, we compare the $|\Tr [\hat{O}(t_n) \hat{O}^\dagger]|$ evaluated with bond dimension $D=400$ and $D=600$. 
For the clean systems, the results of different bond dimensions start to deviate when $t_n\approx 10$ for all system sizes, which indicates the bond dimension is saturated. Therefore we only reserve the simulation result before $t_n=10$. While for the disordered system, the difference between different bond dimensions is not significant, indicating slow entanglement growth. For efficiency consideration, we cut off the simulation at $t_n=20$. 

Now let's estimate the time truncation error. For the clean systems, we assume that $|\Tr [\hat{O}(t_n) \hat{O}^\dagger]|$ after the truncation time $t_n=10$ is smaller than $10^{-6} \times 2^N$, as inferred from Fig. \ref{fig:error}. Then according to Eq. \eqref{eq:epsilonA} and \eqref{eq:txsigma}, we have
\begin{align}
    \epsilon_A \leq \frac{5} {2\pi x x_\omega }(e^{-x^2/2}\times 10^{-5}+e^{-x_\omega^2/2}\times 10^{-6})\times 2^N ,
\end{align}
This means the relative error is small compared to $ A(E,\omega) \approx \Dos(E) S'^{\rho_\sigma(E)}_O(\omega)$, for any $x,x_\omega \sim 1$. 

While for the disordered system, $\Tr [\hat{O}(t_n) \hat{O}^\dagger]$ decays much more slowly. From the Fig. \ref{fig:error} we assume the late-time data is smaller than $10^{-1}$, which gives an error of 
\begin{align}
    \epsilon_A \leq \frac{5} {\pi x x_\omega } (e^{-x^2/2}\times 10^{-5}+e^{-x_\omega^2/2}\times 10^{-1})\times 2^N,
\end{align}
To ensure the smallness of the error, we choose $x_\omega \ge 3$ to suppress the error, 
which, if we bound the simulation time to $t_{\max}\leq 20$, means that we can reach widths 
$\sigma_\omega \ge 0.3$. 

\subsection{Other filter methods}
\label{app:more_filters}
There are various ways to probe the off-diagonal matrix elements using the filter operators. For example, one could use two filters, with one selecting the mean energy and the other acting on the energy difference,
\begin{equation}
    \begin{aligned}
    ( O | \hat{P}^a _{\sigma_E} (E) \hat{P}^c _{\sigma_\omega} (\omega) | O )
     \approx  e^{S(E+\frac{\omega}{2})+S(E-\frac{\omega}{2}) } \overline{|O_{E-\frac{\omega}{2},E+\frac{\omega}{2}}|^2},
\end{aligned}
\end{equation}
or apply two filters at different energies
\begin{equation}
\begin{aligned}
    \Tr[\hat{P}_{\sigma_1}(E_1)\hat{O}\hat{P}_{\sigma_2}(E_2)\hat{O}^\dagger] \approx & e^{S(E_1)+S(E_2) } \overline{|O_{E_1,E_2}|^2}.
\end{aligned}
\end{equation}
If $E_1=E_2$, this quantity equals the two-point regularized correlator defined in \cite{pappalardi2024windows}.

\section{Mathematical details}
\label{app:sf}

\subsection{A Rigorous Expression of $S_{O}^{\rho_\sigma(E)}(\omega)$}
In this appendix, we are going to derive a rigorous expression for $S_O^{\rho_\sigma(E)}(\omega)$ under the condition that ETH is valid. The influence of the finite filter width will be examined in detail. We start with the spectral function for the single eigenstate $\ket{\alpha}$. 
\begin{align}
    S^{\ket{\alpha}}_O(\omega) = \sum_\beta |O_{\alpha\beta}|^2 \delta(\omega-E_\beta+E_\alpha),
\end{align}
If the observable $\hat{O}$ fulfills ETH, one can replace $O_{\alpha\beta}$ with its ETH prediction 
\begin{align}
    S^{\ket{\alpha}}_O(\omega) = \sum_{\beta} e^{-\frac{S(E_\alpha)+S(E_\beta)}{2}} \left| f\left( \frac{E_\alpha+E_\beta}{2},E_\beta-E_\alpha \right)\right|^2 |R_{\alpha\beta}|^2 \delta (\omega-E_\beta+E_\alpha).
\end{align}
For large systems, the eigenenergy spacing is exponentially small, therefore we could substitute $\sum_\beta$ with $\int d E_\beta e^{S(E_\beta)} = \int d\omega' e^{S(E_\alpha+ \omega')}$, and $|R_{\alpha\beta}|^2$ with its variance 1,

\begin{equation}
    \begin{aligned}
    S^\alpha_O(\omega)  =  & \int d \omega' e^{\frac{S(E_\alpha+\omega')-S(E_\alpha)}{2}} \left| f_O(E_\alpha+\omega'/2,\omega')\right|^2 \delta(\omega-\omega') 
    \\ &=  e^{\frac{S(E_\alpha+\omega)-S(E_\alpha)}{2}} \left| f_O(E_\alpha+\omega/2,\omega)\right|^2 \\ & \equiv G(E_\alpha,\omega)
     \end{aligned}
\end{equation}
where $G(E_\alpha,\omega)$ is introduced for simplification. We then proceed to the filter ensemble. The mean energy and the energy variance of the filter ensemble are \cite{Yang2022}
\begin{align}
    \bar{E} = \frac{E}{ 1+ \frac{\sigma^2}{N\sigma_0^2}}, \Delta E  = \frac{\sigma}{\sqrt{ 1+ \frac{\sigma^2}{N\sigma_0^2}}}.
    \label{eq:average&uncertainty}
\end{align}
where $\sigma_0$ is the constant determining the width of $\Dos(E)$ in Eq. \ref{equ:dos}. With that one could obtain
\begin{equation}
\begin{aligned}
    S_O^{\rho_\sigma(E)}(E,\omega) = & \sum_\alpha \bra{\alpha} p_{\sigma}(E) \ket{\alpha} S^{\ket{\alpha}}_O(\omega)
    \\ = & G(\bar{E},\omega) +  \mathcal{O}(\sigma^2) \partial_E^2 G(\bar{E},\omega) .
\end{aligned}
\end{equation}
Let's estimate the scale of the correction. Notice that
\begin{gather}
    \partial^2 G = G \left[  (\partial \ln G)^2 + \partial^2 \ln G\right] 
\end{gather}
\begin{equation}
    \begin{aligned}
    \ln G(\bar{E},\omega) & = \frac{S(\bar{E} + \omega ) - S(\bar{E})}{2} + \ln \left| f_O(\bar{E}+\omega/2,\omega)\right|^2 
    \\ & = \frac{\omega}{2} \beta(\bar{E}) +  \frac{\omega^2}{4}\partial_E \beta(\bar{E}) + \ln \left| f_O(\bar{E}+\omega/2,\omega)\right|^2 +\mathcal{O}\left(\frac{1}{N^2}\right)
    \end{aligned}
\end{equation}
where $\beta(E) = \partial_E S(E)$ is the inverse temperature at the energy $E$ in the microcanonical ensemble. 
Each derivative of $\beta(E)$ and $\left| f_O(E+\omega/2,\omega)\right|^2$ with respect to the extensive quantity $E$ contributes a factor of $O(1/N)$. Therefore,
\begin{align}
    \partial_E \ln G = O(1/N),~\partial_E^2 \ln G = O(1/N^2).
\end{align}
Combining all the derivatives, one obtains
\begin{align}
    S_O^{\rho_\sigma(E)}(\omega) =G(\bar{E},\omega) \left[ 1+ \mathcal{O}\left( \frac{\sigma^2}{N^2} \right)\right].
    \label{eq:S_barE}
\end{align}
One can in addition replace $\bar{E}$ with its expression in Eq. \eqref{eq:average&uncertainty},
\begin{align}
    G(\bar{E},\omega) = G(E,\omega) + \mathcal{O}\left(\frac{\sigma^2}{N}\right)\partial_E G(E,\omega) = G(E,\omega) \left[ 1+ \mathcal{O}\left( \frac{\sigma^2}{N^2} \right)\right].
\end{align}
Altogether we achieve
\begin{align}
    S_O^{\rho_\sigma(E)}(\omega) = e^{\frac{S({E}+\omega)-S({E})}{2}}
 \left| f_O({E}+\omega/2,\omega)\right|^2 \left[ 1+ \mathcal{O}\left( \frac{\sigma^2}{N^2} \right)\right].
\end{align}

\subsection{FDT}
The indicator function of FDT defined in the mean text is
\begin{align}
    \mathrm{\beta}_{FDT}^{\rho_\sigma(E)}:= \frac{1}{\omega} \ln\left[\frac{S^{\rho_\sigma(E)}_O(\omega)}{S^{\rho_\sigma(E)}_O(-\omega)}\right].
\end{align}
To obtain the indicator function, we expand $\ln S^{\rho_\sigma(E)}_O(\omega)$ in Eq. \ref{eq:S_barE} around $\omega = 0$,
\begin{equation}
    \begin{aligned}
    \ln S_{O}^{\rho_\sigma(E)}(\pm\omega) = &\pm\frac{\beta(\bar{E}) \omega}{2} + \frac{\partial \beta(\bar{E})}{\partial E} \frac{\omega^2}{4} + \mathcal{O}\left(\frac{1}{N^2}\right)+ 
     \ln \left|f_O(\bar{E},\omega)\right|^2
     \pm\frac{\omega}{2}\partial_E \ln \left|f_O(\bar{E},\omega)\right|^2 +\mathcal{O}\left(\frac{\sigma^2}{N^2}\right).
    \end{aligned}
\end{equation}
Notice that the terms with an even power of $\omega$ cancel out in the indicator function,
\begin{equation}
    \mathrm{\beta}_{FDT}^{\rho_\sigma(E)} = \beta(\bar{E}) + \partial_E \ln \left|f_O(\bar{E},\omega)\right|^2 + \mathcal{O}\left(\frac{\sigma^2}{N^2}\right)+ \mathcal{O}\left(\frac{1}{N^2}\right).
\end{equation}
The first correction term $\partial_E \ln \left|f_O(\bar{E},\omega)\right|^2 $ scales as $\mathcal{O}(1/N)$. As the ensemble width scales as $\sigma = \mathcal{O}(\sqrt{N})$ in our simulation, the second correction term is also $\mathcal{O}(1/N)$. All the corrections vanish in the thermodynamic limit, meaning that $\mathrm{\beta}_{FDT}^{\rho_\sigma(E)}$ converges to the thermal $\beta$ in the thermodynamic limit. Notice that a similar derivation was in \cite{numFDT}, which is in line with our result.

\subsection{The generalized spectral function}
In this section, we are going to analyze the numerical error of replacing the $\delta$ function in the spectral function with a Gaussian filter with width $\sigma_\omega$. The generalized spectral function for an ensemble $\rho$ can be written as
\begin{equation}
\begin{aligned}
    S'^{\rho}_O(\omega) = &\sum_{\alpha\beta} \bra{\alpha}\hat{\rho}\ket{\alpha}|O_{\alpha\beta}|  g_{\sigma_\omega}(\omega-E_\beta+E_\alpha)
    \\= & \sum_{\alpha\beta} \bra{\alpha}\hat{\rho}\ket{\alpha}|O_{\alpha\beta}|   \int d\omega' g_{\sigma_\omega}(\omega-\omega') \delta(\omega'-E_\beta+E_\alpha)
    \\ = & \int d\omega' g_{\sigma_\omega}(\omega-\omega') S^{\rho}_O(\omega').
    \label{equ:s'_s}
\end{aligned}
\end{equation}
Eq. \eqref{equ:s'_s} gives another interpretation of  $S'^{\rho}_O(\omega)$: it is a convolution of $S^{\rho}_O(\omega)$ with a filter. Depending on the specific $S^{\rho}_O(\omega)$, this convolution will give different errors. 

1. If $S^{\rho}_O(\omega)$ is a continues function of $\omega$, which means it fulfills the following condition
\begin{align}
    \left|S^{\rho}_O(\omega+\Delta \omega)- S^{\rho}_O(\omega) \right| \leq  K \left|\Delta \omega \right|
\end{align}
where $K$ is a positive constant. Then we have
\begin{align}
    S'^{\rho}_O(\omega) \leq \int d\omega' g_{\sigma_\omega}(\omega-\omega') \left[S^{\rho}_O(\omega)+K|\omega'-\omega|\right] = S^{\rho}_O(\omega) + K \mathcal{O}(\sigma_\omega).
\end{align}
The error is of the order $\mathcal{O}(\sigma_\omega)$.

2. For the observable $O$ fulfilling ETH, $S^{\rho}_O(\omega)$ should be a smooth function of $\omega$. Using the property of the Gaussian function, we have
\begin{align}
    S'^{\rho}_O(\omega) = S_O^{\rho} (\omega) +\frac{1}{2} \sigma_\omega^2 \partial_\omega ^2 S_O^{\rho} (\omega) +\mathcal{O}(\sigma_\omega^4)
\end{align}
The relative error is of the order $\mathcal{O}(\sigma_\omega^2)$. 
In general, we expect the smoother $S^{\rho}_O(\omega)$ is, the smaller this error would be.

\section{Solve the Integrable Ising model}
\label{app:JW}
When $J_2,r=0$, the model in Eq. \ref{equ:ham} becomes
\begin{align}
    \hat{H}= -J \sum_{i=1}^{N-1} {\sigma}_i^z{\sigma}_{i+1}^z - g\sum_{i=1}^{N}\hat{\sigma}_i^x.
\end{align}
It is well known that we could rewrite the spin Hamiltonian with fermion creation operators by Jordan-Wigner Transformation, where the correspondence of spin operators and the fermion operators are defined as
\begin{gather}
    {\sigma}_j^z = (-1)^{\sum_{k<j}\hat{n}_k}\left(\hat{c}_j^\dagger + \hat{c}_j\right),
    \\
    {\sigma}_j^y = -i(-1)^{\sum_{k<j}\hat{n}_k}\left(\hat{c}_j^\dagger - \hat{c}_j\right),
    \\
    {\sigma}_j^x = \hat{c}_j\hat{c}_j^\dagger -\hat{c}_j^\dagger\hat{c}_j.
\end{gather}
Applying this transformation to the original spin Hamiltonian, we can get an equivalent fermionic Hamiltonian
\begin{align}
    \hat{H}= - J \sum_{i=1}^{N-1}\left(\hat{c}_i^\dagger\hat{c}_{i+1}+\hat{c}_i^\dagger\hat{c}_{i+1}^{\dagger}\right)  + g \sum_{i=1}^{N} \left(\hat{c}_i^\dagger\hat{c}_i - \hat{c}_i\hat{c}_i^\dagger  \right).
\end{align}
This Hamiltonian is quadratic in fermionic operators and thus can be diagonalized using a Bogoliubov transformation
\begin{align}
    \hat{c}_i = \sum_{\mu} u_{i\mu} \hat{\gamma}_\mu + v_{i\mu} \hat{\gamma}_\mu^\dagger
\end{align}
where $\hat{\gamma}_\mu,\hat{\gamma}_\mu^\dagger$ are bogoliubov fermions. The ground state is the state annihilated by all $\hat{\gamma}_\mu$, which we denote by $\ket{\emptyset_\gamma}$. The eigenstates are fock states of $\hat{\gamma}_\mu$,
\begin{gather}
    \ket{\{n_\mu\}} = \prod _{\mu=1}^{L}\left(\gamma_\mu^\dagger \right)^{n_\mu}\ket{\emptyset_\gamma},\quad with \quad n_\mu = 0,1
    \\
    E_{\{n_\mu\}} = \sum_\mu (2 n_\mu-1) \epsilon_\mu.
\end{gather}

This Hamiltonian features an inherent particle-hole symmetry: for any given eigenstate $\ket{\{n_\mu\}}$, if we define $\mathcal{S}$ as the operation that interchanges particles and holes, then $\mathcal{S}\ket{\{n_\mu\}}$ remains an eigenstate with an energy of the opposite sign. This results in the symmetry in the matrix elements of the observable $\hat{O}=\sigma_{N/2}^z$. To be concrete, one could check that the matrix elements of $\sigma_{N/2}^z$ satisfy
\begin{align}
    \left|\bra{\{m_\nu\}} \sigma_{N/2}^z \ket{\{n_\mu\} }\right| = \left|\bra{\{m_\nu\}} \left( \hat{c}_{N/2}+\hat{c}_{N/2}^\dagger \right)\ket{\{n_\mu\} }\right| =\left|\bra{ \mathcal{S}  \{m_\nu\}} \left( \hat{c}_{N/2}+\hat{c}_{N/2}^\dagger \right)\ket{ \mathcal{S}  \{n_\mu\}}\right|.
\end{align}
This means for every matrix element, there is another matrix element with equal weight at opposite energies. As $V_O(E,\omega)$ is an average over matrix elements
\begin{align}
    V_O(E,\omega) = e^{\frac{S(E-\omega/2)+S(E+\omega/2)}{2}}\overline{|O_{E-\omega/2,E+\omega/2}|^2}.
\end{align}
The symmetry in matrix elements immediately implies $V_O(E,\omega)= V_O(-E,-\omega)$.

\end{document}